\documentclass[jcp,aip,amsmath,amssymb,reprint,USenglish]{revtex4-2}

\usepackage{graphicx}
\usepackage{dcolumn}
\usepackage{bm}
\usepackage[utf8]{inputenc}
\usepackage[T1]{fontenc}
\usepackage{mathptmx}
\usepackage{etoolbox}
\usepackage{physics}
\usepackage[colorlinks=true, allcolors=blue]{hyperref}
\usepackage[normalem]{ulem} 
\usepackage{xcolor} 
\usepackage{ulem}
\usepackage{soul}
\usepackage{mathtools}

\DeclareMathAlphabet\mathbfcal{OMS}{cmsy}{b}{n}

\renewcommand{\Re}{\operatorname{Re}}
\renewcommand{\Im}{\operatorname{Im}}

\begin{document}

\def\equationautorefname#1#2\null{Eq.#1(#2\null)}
\renewcommand{\figureautorefname}{Fig.}

\preprint{AIP/123-QED}

\title{General theory of cavity-mediated interactions between low-energy matter excitations}
\author{Carlos J. Sánchez Martínez}
\author{Frieder Lindel}
\author{Francisco J. García-Vidal}
\author{Johannes Feist}
\email[Corresponding author:\;]{johannes.feist@uam.es}
\affiliation{Departamento de Física Teórica de la Materia Condensada, Universidad Autónoma de Madrid, E-28049 Madrid, Spain}
\affiliation{Condensed Matter Physics Center (IFIMAC), Universidad Autónoma de Madrid, E-28049 Madrid, Spain}


\begin{abstract}
The manipulation of low-energy matter properties such as superconductivity, ferromagnetism and ferroelectricity via cavity quantum electrodynamics engineering has been suggested as a way to enhance these many-body collective phenomena. In this work, we investigate the effective interactions between low-energy matter excitations induced by the off-resonant coupling with cavity electromagnetic modes.  We extend previous work by going beyond the dipole approximation accounting for the full polarization and magnetization densities of matter. We further include the often neglected diamagnetic interaction and, for the cavity, we consider general linear absorbing media with possibly non-local and non-reciprocal response. We demonstrate that, even in this general scenario, the effective cavity-induced interactions between the matter degrees of freedom are of electrostatic and magnetostatic nature. This confirms the necessity of a multimode description for cavity engineering of matter systems where the low-energy assumption holds. Our findings provide a theoretical framework for studying the influence of general optical environments on extended low-energy matter excitations.
\end{abstract}


\maketitle 

\section{Introduction}

The interaction between light and matter has been an intense field of research since the foundations of quantum electrodynamics (QED) in the last century. One of the most prominent results in QED is the phenomenon of spontaneous emission~\cite{Dirac1927}: an emitter in an excited state is coupled to the quantum fluctuations of the electromagnetic (EM) vacuum, allowing it to transfer its energy to an unoccupied EM state through the emission of a photon. Vacuum EM fields also induce an energy shift on the emitter energy levels~\cite{Bethe1947}. However, in free space, these two effects are relatively weak. A way to enhance light-matter interactions is to place the emitter inside a cavity, which confines the EM field and thus increases the probability of interaction between photons and the emitter. A widely-known result of this strategy is the increase of the spontaneous emission rate, predicted by Purcell in $1946$~\cite{Purcell1946}. This prediction gave birth to the field of cavity QED, opening a new paradigm to control light-matter interactions through cavity engineering of the vacuum EM field.

In an early stage, material structures such as Fabry-Pérot cavities filled with emitters were used to obtain new light sources, with perhaps the best example being the laser~\cite{Schawlow1958}. Although this field of research remains active, especially with the advent of quantum light sources~\cite{Shields2007}, in recent years the inverse approach of modifying matter properties via cavity-enhanced vacuum fields~\cite{GarciaVidal2021,Schlawin2022} has attracted increasing interest. Two regimes are typically explored: on the one hand, the regime where the matter transition frequencies are resonant with a cavity mode; on the other hand, the off-resonant regime where low-energy matter excitations are considered.

The resonant regime has attracted much attention in the community since, if the light-matter interaction is strong enough, hybrid light-matter states called polaritons are formed, with properties different from those of the individual bare matter and EM states. Both theoretical and experimental investigations have been carried out to examine these new properties. A striking example is the condensation of polaritons due to their (approximately) bosonic nature and effective interactions mediated by the material component~\cite{Kasprzak2006, Daskalakis2014, Ramezani2017Plasmon}. Later studies showed modifications of chemical reactions~\cite{Hutchison2012, Nagarajan2021}, enhancement of exciton transport~\cite{JFeist2015,JSchachenmayer2015,Lerario2017,Shaocong2020} and energy transfer in organic materials~\cite{Coles2014,Groenhof2019}, or the control of topological phases of matter~\cite{Downing2019}. In addition, the cavity-mediated magneto-transport of two-dimensional electron gases has also been studied, showing that the linear direct-current resistivity is substantially modified~\cite{ParaviciniBagliani2019} as well as the integer and fractional quantum Hall regimes~\cite{Appugliese2022,Enker2024}.  

However, there is still a dearth of experimental results for the cavity-modification of low-energy matter properties. In this case, the light-matter interaction mechanism is completely different from polariton physics. As the matter excitations are detuned from the cavity EM field, the interaction is off-resonant, resulting in an cavity-induced effective matter-matter interaction mediated by virtual photons. Note that these cavity-induced interactions do not necessarily have to be directly mediated by the cavity field. It is also possible that additional degrees of freedom such as phonons, plasmons or magnons are coupled to both the matter excitations and the cavity EM modes, so that in this case the EM modes indirectly mediate the effective matter-matter interactions. Experimentally, an apparent cavity-induced change in the temperature of a metal-insulator phase transition has been found, although the most likely explanation seems to be that this is due to radiative heat transfer~\cite{Jarc2023}, which is a well-established effect that does not rely on microscopic modifications of material properties~\cite{Kim2015Radiative}. Theoretical works in the off-resonant regime have predicted modifications in superconductivity~\cite{Sentef2018,Schlawin2019,Curtis2019,Hagenmuller2019,Klinovaja2024,Polini2024}, ferromagnetism~\cite{Roche2021,Thomas2021}, ferroelectricity~\cite{Ashida2020,Latini2021,Lenk2022}, Moiré materials~\cite{Jiang2024,Masuki2024}, and Fermi liquids~\cite{Riolo2024}. 
Recently, a theoretical work~\cite{Pantazopoulos2024} has addressed this question from a general perspective. Under the electric-dipole approximation for the light-matter interaction, this work demonstrated that the cavity-induced effective interactions between low-energy matter excitations, induced either directly or indirectly by the cavity EM modes, are described by the EM Green's tensor evaluated at zero frequency and are thus electrostatic in nature~\cite{Barcellona2018}. The main assumption in this work is the consideration of low-energy matter excitations, with frequencies much smaller than that of any cavity mode, thus obtaining a fully non-resonant effective interaction that implies the necessity of summing over all the cavity EM modes (although it should be noted that in the thermodynamic limit, the same result is obtained even without an explicit low-frequency approximation as long as there is no resonant contribution~\cite{Roman-Roche2022}). When this assumption is applicable, this theoretical framework shows that cavity engineering can be easily understood through the conceptually simple consideration of electrostatic interactions, which on the other hand also implies that the possibilities of cavity-induced manipulation of materials are somewhat limited. Nevertheless, this formalism has been applied to demonstrate theoretically the emergence of a biquadratic long-range interaction between spins mediated by their coupling to phonons hybridized with vacuum photons into polaritons~\cite{Pantazopoulos2024_2}.

In the present article, we provide an extension of the previous work~\cite{Pantazopoulos2024} for the case where the matter states are directly coupled to the cavity EM modes. We go beyond the electric dipole approximation for the light-matter interaction, considering the full polarization and magnetization densities of the matter elements, as well as the often neglected diamagnetic term. The polarization and magnetization densities include the electric and magnetic interactions to all multipole orders, respectively. This extension is particularly relevant when considering extended low-energy matter excitations or nanophotonic platforms~\cite{GonzalezTudela2024} where subwavelength field confinement is achieved and strong field gradients can challenge the validity of the dipole approximation~\cite{Huang2024}. Furthermore, we allow the ``cavity'' structure to be made out of general linear absorbing media, including non-local and non-reciprocal media, via the theoretical framework of macroscopic QED~\cite{Scheel2008,Buhmann2012BI,Huttner1992}. We derive an effective Hamiltonian for the matter system by using a coherent-path-integral formulation of the partition function in the same way than in Ref.~\citenum{Pantazopoulos2024}. We demonstrate that also with these generalizations, we recover the zero-frequency nature of the off-resonant cavity-induced low-energy matter-matter interactions. Importantly, this result emerges only when all field modes are considered, highlighting that models relying on single modes can lead to misleading results.

\section{Effective Hamiltonian}

\subsection{General mode description}

We start by defining the Hamiltonian for our composite system consisting of low-energy matter excitations coupled to the EM field supported by arbitrary cavity structures. We consider a matter system, described by $N$ polarization and magnetization densities (each one describing a set of degrees of freedom), interacting with a set of quantized EM modes $\left\lbrace \hat{a}_n \right\rbrace$ in the Power-Zienau-Woolley (PZW) picture~\cite{Power1959}. The Hamiltonian can be written as~\cite{Buhmann2012BI}
\begin{multline}\label{eq:H}
    \hat{H}=\hat{H}_{le}+\sum_n \hbar \omega_n \hat{a}^{\dagger}_n \hat{a}_n - \int d^3r \sum_i^N \hat{\mathbf{P}}_i\left(\mathbf{r}\right)\cdot \hat{\mathbf{E}}\left(\mathbf{r}\right) 
    \\
    - \int d^3r \sum_i^N \hat{\mathbf{M}}_i\left(\mathbf{r}\right)\cdot \hat{\mathbf{B}}\left(\mathbf{r}\right) + \sum_i^N \sum_{\gamma \in i} \left[\int d^3r \hspace{0.05cm}\hat{\boldsymbol{\Theta}}_{\gamma}\left(\mathbf{r}\right)\times \hat{\mathbf{B}}\left(\mathbf{r}\right)  \right]^2,
\end{multline}
where $\hat{H}_{le}$ is the bare low-energy matter Hamiltonian, for which we assume that all internal energies are negligible compared to the relevant EM mode frequencies. The matter Hamiltonian includes the full information about the matter excitations, the direct interaction between them and the dipole self-energy term. The bare EM Hamiltonian is represented by a sum over modes with frequencies $\omega_n$. The linear interacting terms are written in terms of the electric and magnetic field operators $\hat{\mathbf{E}}\left(\mathbf{r}\right)=\sum_n\hat{\mathbf{E}}_n\left(\mathbf{r}\right)=\sum_n\left[\mathbf{E}_n\left(\mathbf{r} \right)\hat{a}_n + \mathbf{E}^{*}_n\left(\mathbf{r} \right)\hat{a}^{\dagger}_n\right]$ and $\hat{\mathbf{B}}\left(\mathbf{r}\right)=\sum_n\hat{\mathbf{B}}_n\left(\mathbf{r}\right)=\sum_n\left[\mathbf{B}_n\left(\mathbf{r} \right)\hat{a}_n + \mathbf{B}^{*}_n\left(\mathbf{r} \right)\hat{a}^{\dagger}_n\right]$, where the relation between the coefficients $\mathbf{E}_n(\mathbf{r})$ and $\mathbf{B}_n(\mathbf{r})$ is given by $\mathbf{B}_n(\mathbf{r})=\frac{1}{i\omega_n}\nabla\times\mathbf{E}_n(\mathbf{r})$. The Hermitian operators, $\hat{\mathbf{P}}_i\left(\mathbf{r}\right)$ and $\hat{\mathbf{M}}_i\left(\mathbf{r}\right)$, represent the polarization and generalized magnetization densities of the i-th matter element, respectively. The generalized magnetization density includes both the standard atomic magnetization and the so-called Röntgen magnetization~\cite{Baxter1993} given by the centre-of-mass motion of the i-th matter element. Both polarization and magnetization density operators can be formally defined as sums of line integrals of delta functions along curves joining an arbitrary reference point to the positions of the charges composing each i-th matter element~\cite{Healy1978}. The last term in \autoref{eq:H} is the nonlinear diamagnetic term, where the Hermitian density operator $\hat{\boldsymbol{\Theta}}_{\gamma}\left(\mathbf{r}\right)$ depends exclusively on the internal structure (represented by the index $\gamma$) of the i-th matter element~\cite{Buhmann2012BI}. In particular, it is a function of only the position operator of both the i-th matter element and the individual components of that i-th matter element. 

We are interested in obtaining an effective Hamiltonian for the dynamics of the matter degrees of freedom by tracing out the EM field. The thermodynamic properties of the matter degrees of freedom are encoded in the canonical partition function, $Z=\textup{Tr}_{\textup{M+F}}\left[\exp\left(-\beta\hat{H}\right)\right]$, with $\beta=(k_B T)^{-1}$ and the Boltzmann constant $k_B$. The subscript $\textup{M+F}$ in the trace indicates that the trace is performed over all matter and field degrees of freedom. Leveraging the definition of the partition function, we can utilize a coherent-state path integral approach~\cite{Coleman2015,Roman-Roche2022} to trace out the EM environment and obtain an effective matter Hamiltonian:
\begin{equation}\label{eq:Z}
    Z=\textup{Tr}_{\textup{M}}\left[\exp\left(-\beta\hat{H}_{\textup{eff}}\right)\right],
\end{equation}
with
\begin{equation}
    \exp\left(-\beta\hat{H}_{\textup{eff}}\right)=\prod_n\left(\int\frac{d^2\alpha_n}{\pi}\right)\bra{\alpha}\exp\left(-\beta\hat{H}\right)\ket{\alpha}.
\end{equation}
Here, $\ket{\alpha}=\bigotimes_n \ket{\alpha_n}=\ket{\alpha_1}\otimes \ket{\alpha_2}\otimes \ldots$ is the tensor product of the coherent states associated with each EM mode, defined by $\hat{a}_n\ket{\alpha_n}=\alpha_n\ket{\alpha_n}$. After some algebra, detailed in Appendix~\ref{app:A}, it can be shown that the effective Hamiltonian reads as
\begin{multline}\label{eq:Heff}
    \hat{H}_{\textup{eff}}=\hat{\tilde{H}}_{le} - \sum_{i,j}\int d^3r\int d^3r^{\prime}\left[\hat{\mathbf{P}}_i\left(\mathbf{r}\right)\cdot \boldsymbol{\lambda}^{ee}\left(\mathbf{r},\mathbf{r'}\right)\cdot\hat{\mathbf{P}}_j\left(\mathbf{r'}\right) \right.
    \\
    \left. + \hat{\mathbf{P}}_i\left(\mathbf{r}\right)\cdot \boldsymbol{\lambda}^{em}\left(\mathbf{r},\mathbf{r'}\right)\cdot\hat{\mathbf{M}}_j\left(\mathbf{r'}\right) \right.
    \\
    \left. + \hat{\mathbf{M}}_i\left(\mathbf{r}\right)\cdot \boldsymbol{\lambda}^{me}\left(\mathbf{r},\mathbf{r'}\right)\cdot\hat{\mathbf{P}}_j\left(\mathbf{r'}\right) \right.
    \\
    \left.+ \hat{\mathbf{M}}_i\left(\mathbf{r}\right)\cdot \boldsymbol{\lambda}^{mm}\left(\mathbf{r},\mathbf{r'}\right)\cdot\hat{\mathbf{M}}_j\left(\mathbf{r'}\right)\right],
\end{multline}
where $\hat{\tilde{H}}_{le}=\hat{H}_{le}+\hat{H}_{ren}$ is the bare low-energy matter Hamiltonian corrected by a renormalization contribution arising from the diamagnetic term in \autoref{eq:H}:
\begin{equation}\label{eq:Hrenorm}
    \hat{H}_{ren} = \sum_i\sum_{\gamma\in i}\int d^3r\int d^3r^{\prime}\textup{Tr}\left[\hat{\boldsymbol{\Theta}}_{\gamma}\left(\mathbf{r}\right)\times\boldsymbol{\Omega}\left(\mathbf{r},\mathbf{r'}\right)\times\hat{\boldsymbol{\Theta}}_{\gamma}\left(\mathbf{r'}\right)\right].
\end{equation}
The diamagnetic tensor field $\boldsymbol{\Omega}\left(\mathbf{r},\mathbf{r'}\right)$ is defined as:
\begin{equation}\label{eq:Omega_tilde}
    \boldsymbol{\Omega}\left(\mathbf{r},\mathbf{r'}\right) = \overrightarrow\nabla_r\times\left\lbrace \sum_n\mathrm{Re}\left[\frac{\mathbf{E}_n\left(\mathbf{r}\right)\otimes\mathbf{E}_n^{*}\left(\mathbf{r'}\right)}{\omega_n^2}\right]\right\rbrace\times\overleftarrow\nabla_{r'}.
\end{equation}
Here, we have used a compact notation for the two-side curl of a tensor field $\mathbf{T}(\mathbf{r},\mathbf{r'})$: $\left[\overrightarrow\nabla_r\times\mathbf{T}(\mathbf{r},\mathbf{r'})\times\overleftarrow\nabla_{r'}\right]^{\alpha\beta} = \sum_{\gamma\delta\mu\nu}\varepsilon^{\alpha\gamma\delta}\varepsilon^{\beta\mu\nu}\partial^{\gamma}_r\partial^{\mu}_{r'}T^{\delta\nu}(\mathbf{r},\mathbf{r'})$. In the same way, the left- and right-side curl are defined as $\left[\overrightarrow\nabla_r\times\mathbf{T}(\mathbf{r},\mathbf{r'})\right]^{\alpha\beta} = \sum_{\gamma\delta}\varepsilon^{\alpha\gamma\delta}\partial^{\gamma}_r T^{\delta\beta}(\mathbf{r},\mathbf{r'})$ and $\left[\mathbf{T}(\mathbf{r},\mathbf{r'})\times\overleftarrow\nabla_{r'}\right]^{\alpha\beta} = \sum_{\gamma\delta} \varepsilon^{\beta\gamma\delta}\partial^{\gamma}_{r'}T^{\alpha\delta}(\mathbf{r},\mathbf{r'})$, respectively.

It can be shown that $\boldsymbol{\Omega}\left(\mathbf{r},\mathbf{r'}\right)$ is equivalent to the equal-time correlation function of the magnetic field over the EM vacuum state $\ket{0}$: $\boldsymbol{\Omega}\left(\mathbf{r},\mathbf{r'}\right)=\mathrm{Re}[\left\langle 0| \hat{\mathbf{B}}\left(\mathbf{r}\right)\otimes\hat{\mathbf{B}}\left(\mathbf{r'}\right)\right |0\rangle]$. This shows that the diamagnetic renormalisation results from correlations in the vacuum magnetic cavity field. Another important feature on the nature of the diamagnetic renormalization in \autoref{eq:Omega_tilde} is that the emission and reabsorption of the virtual photons responsible of the interaction occur in the same matter element. This is in contrast to the rest of terms on the right-hand side of \autoref{eq:Heff}, which account for the cavity-mediated interactions between the polarization and magnetization densities of the different matter elements. In total there are four different effective couplings: a coupling between polarization densities $\boldsymbol{\lambda}^{ee}\left(\mathbf{r},\mathbf{r'}\right)$, a coupling between magnetization densities $\boldsymbol{\lambda}^{mm}\left(\mathbf{r},\mathbf{r'}\right)$, and two cross-couplings $\boldsymbol{\lambda}^{em}\left(\mathbf{r},\mathbf{r'}\right)$ and $\boldsymbol{\lambda}^{me}\left(\mathbf{r},\mathbf{r'}\right)$. They are given by:
\begin{subequations}\label{eq:effcoupling}
\begin{align}
\boldsymbol{\lambda}^{ee}\left(\mathbf{r},\mathbf{r'}\right) &= \sum_n \Re \left[\frac{\mathbf{E}_n\left( \mathbf{r} \right)\otimes \mathbf{E}^{*}_n\left(\mathbf{r'} \right)}{\hbar \omega_n} \right], \\
\boldsymbol{\lambda}^{em}\left(\mathbf{r},\mathbf{r'}\right) &= - \sum_n \left\lbrace \Im\left[\frac{\mathbf{E}_n\left( \mathbf{r} \right)\otimes \mathbf{E}^{*}_n\left(\mathbf{r'} \right)}{\hbar \omega_n^2} \right] \times\overleftarrow\nabla_{r'}\right\rbrace, \\
\boldsymbol{\lambda}^{me}\left(\mathbf{r},\mathbf{r'}\right) &= \sum_n \left\lbrace \overrightarrow\nabla_r\times\Im\left[\frac{\mathbf{E}_n\left( \mathbf{r} \right)\otimes \mathbf{E}^{*}_n\left(\mathbf{r'} \right)}{\hbar \omega_n^2} \right]\right\rbrace, \\
\boldsymbol{\lambda}^{mm}\left(\mathbf{r},\mathbf{r'}\right) &= \sum_n \left\lbrace\overrightarrow\nabla_r\times\Re\left[\frac{\mathbf{E}_n\left( \mathbf{r} \right)\otimes\mathbf{E}^{*}_n\left(\mathbf{r'}\right)}{\hbar \omega_n^3} \right] \times\overleftarrow\nabla_{r'} \right\rbrace.
\end{align}
\end{subequations}
Note that as a consequence of the low-energy assumption for the matter excitations, these four effective couplings are independent of the energy of the matter states, manifesting the non-resonant nature of the interaction. In addition, the structure of the couplings is independent of the level of approximation made regarding the polarization and magnetization densities. In comparison with the results obtained in the dipole approximation~\cite{Pantazopoulos2024}, we find that incorporating the full polarisation and magnetisation fields leads to additional space integrals in the effective Hamiltonian that account for the extended matter excitations [compare~\autoref{eq:Heff}]. The dependence of the effective couplings on the field, however, is still encapsulated in the tensor field 
\begin{equation}\label{eq:em_tensor}
    \mathbfcal{F}_n(\mathbf{r},\mathbf{r'}) \equiv \mathbf{E}_n\left(\mathbf{r}\right)\otimes\mathbf{E}^{*}_n\left(\mathbf{r'}\right),
\end{equation}
whose properties will determine the electromagnetic character of the transitions responsible for the effective matter-matter interactions.

\subsection{Mode summation via the macroscopic QED quantization scheme}

The non-resonant nature of the effective couplings in \autoref{eq:effcoupling} implies that single-mode descriptions of the EM field could be misleading. Instead, the sum over all the modes must be considered. To perform this summation explicitly, we use macroscopic QED~\cite{Huttner1992,Scheel2008,Buhmann2012BI}, which provides a quantization scheme for the EM field in arbitrary dispersive and lossy optical environments with linear response. We note explicitly that we do not rely on the commonly used assumptions of local and reciprocal media. As we show below, the general form of the results does not depend on these assumptions, although some effective coupling terms will be seen to vanish for reciprocal media.

Before performing the sum over all the modes, we briefly summarize the quantization procedure of the EM field given by macroscopic QED in general non-local and non-reciprocal media~\cite{Buhmann2012}: The partial differential equation for the classical electric field resulting from Maxwells equations reads
\begin{multline}\label{eq:dif_equation}
    \overrightarrow\nabla_r \times \overrightarrow\nabla_r \times\mathbf{E}(\mathbf{r},\omega)-\frac{\omega^2}{c^2}\mathbf{E}(\mathbf{r},\omega)
    \\
    -i\mu_0\omega\int d^3r'\hspace{0.02cm}\mathbf{Q}(\mathbf{r},\mathbf{r'},\omega)\cdot\mathbf{E}(\mathbf{r'},\omega)=i\mu_0\omega\mathbf{j}_N(\mathbf{r},\omega),
\end{multline}
where $\mathbf{Q}(\mathbf{r},\mathbf{r'},\omega)$ is the conductivity tensor, which fulfills the Schwarz reflection principle $\mathbf{Q}^{*}(\mathbf{r},\mathbf{r'},\omega)=\mathbf{Q}(\mathbf{r},\mathbf{r'},-\omega^{*})$, $\mathbf{j}_N(\mathbf{r},\omega)$ is the noise current, and the constants $c$ and $\mu_0$ are the vacuum speed of light and the vacuum permeability, respectively (related by $\epsilon_0\mu_0=1/c^2$, with $\epsilon_0$ being the electric constant). The most general form of $\mathbf{Q}(\mathbf{r},\mathbf{r'},\omega)$ in nonlocal and nonreciprocal media can be found in Ref.~\citenum{Buhmann2012} and is used in Appendix~\ref{app:B} to discuss the behavior of the Green's function in the low- and high-frequency limits.

The solution of \autoref{eq:dif_equation} can be formally expressed as
\begin{equation}\label{eq:electricfield}
    \mathbf{E}(\mathbf{r},\omega)=i\mu_0\omega\int d^3r'\hspace{0.02cm}\mathbf{G}(\mathbf{r},\mathbf{r'},\omega)\cdot\mathbf{j}_N(\mathbf{r},\omega),
\end{equation}
where $\mathbf{G}(\mathbf{r},\mathbf{r'},\omega)$ is the classical dyadic EM Green's tensor defined via 
\begin{multline}\label{eq:dif_equationGrenn}
  \left(\overrightarrow\nabla_r \times \overrightarrow\nabla_r \times -\frac{\omega^2}{c^2}\right) \mathbf{G}(\mathbf{r},\mathbf{r'},\omega)
    \\
    -i\mu_0\omega\int d^3 s \hspace{0.05cm}\mathbf{Q}(\mathbf{r},\mathbf{s},\omega)\cdot \mathbf{G}(\mathbf{s},\mathbf{r'},\omega) = \boldsymbol{\delta}(\mathbf{r}-\mathbf{r}^\prime ),
\end{multline}
and $\mathbf{G}(\mathbf{r},\mathbf{r'},\omega) \to \boldsymbol{0}$ for $|\mathbf{r}-\mathbf{r'}| \to \infty$. The quantity $\boldsymbol{\delta}(\mathbf{r}-\mathbf{r'})$ is the delta tensor, defined by $\boldsymbol{\delta}(\mathbf{r}-\mathbf{r'})=\mathbf{I}\delta(\mathbf{r}-\mathbf{r'})$ where $\mathbf{I}$ is the identity tensor and $\delta(\mathbf{r}-\mathbf{r'})$ the delta function.

The EM field can then be quantized by choosing $\mathbf{j}_N(\mathbf{r},\omega)$ as the dynamical variable and promoting it to operator form $\hat{\mathbf{j}}_N(\mathbf{r},\omega)$, with commutation relations $\left[\hat{\mathbf{j}}_N(\mathbf{r},\omega),\hat{\mathbf{j}}^{\dagger}_N(\mathbf{r'},\omega')\right]=\frac{\hbar\omega}{\pi}\delta(\omega-\omega')\mathbf{Q}_{\textup{H}}(\mathbf{r},\mathbf{r'},\omega)$ where $\mathbf{Q}_{\textup{H}}(\mathbf{r},\mathbf{r'},\omega) \equiv [\mathbf{Q}(\mathbf{r},\mathbf{r'},\omega)+\mathbf{Q}^\dagger(\mathbf{r'}, \mathbf{r},\omega)]/2 $ is the Hermitian part of the conductivity tensor. Note that these commutation relations are not bosonic, but bosonic operators $\hat{\mathbf{f}}(\mathbf{r},\omega)$ can be defined by
\begin{equation}\label{eq:bosonic_operator}
 \hat{\mathbf{j}}_N(\mathbf{r},\omega)=\sqrt{\frac{\hbar\omega}{\pi}}\int d^3r'\hspace{0.02cm}\mathbf{R}(\mathbf{r},\mathbf{r'},\omega)\cdot\hat{\mathbf{f}}(\mathbf{r},\omega),
\end{equation}
such that $\left[\hat{\mathbf{f}}(\mathbf{r},\omega),\hat{\mathbf{f}}^{\dagger}(\mathbf{r'},\omega')\right]=\delta(\omega-\omega')\boldsymbol{\delta}(\mathbf{r}-\mathbf{r'})$ where $\mathbf{R}(\mathbf{r},\mathbf{r'},\omega)$ is defined via $\int d^3s\hspace{0.05cm}\mathbf{R}(\mathbf{r},\mathbf{s},\omega)\cdot\mathbf{R}^{\dagger}(\mathbf{r'},\mathbf{s},\omega)=\mathbf{Q}_{\textup{H}}(\mathbf{r},\mathbf{r'},\omega)$. With all this machinery, we can define the electric field amplitude as:
\begin{equation}\label{eq:generalized_green_function}
    \mathbfcal{G}(\mathbf{r},\mathbf{r'},\omega)=i\omega\sqrt{\frac{\hbar\omega}{\pi}}\int d^3s\hspace{0.05cm}\mathbf{G}(\mathbf{r},\mathbf{s},\omega)\cdot\mathbf{R}(\mathbf{s},\mathbf{r'},\omega),
\end{equation}
such that the electric field operator can be written in the following compact way:
\begin{equation}\label{eq:electric_field_operator}
    \hat{\mathbf{E}}(\mathbf{r})=\int_0^{\infty}d\omega\int d^3r'\hspace{0.05cm}\mathbfcal{G}(\mathbf{r},\mathbf{r'},\omega)\cdot\hat{\mathbf{f}}(\mathbf{r'},\omega) + \textup{h.c.}.
\end{equation}

This definition allows us to write the EM tensor field $\mathbfcal{F}_n(\mathbf{r},\mathbf{r'})$ in the macroscopic QED formalism in an elegant way:
\begin{multline}\label{eq:em_tensor_mqed}
    \mathbfcal{F}^{\textup{mQED}}(\mathbf{r},\mathbf{r'},\omega)=\int d^3s\hspace{0.05cm}\mathbfcal{G}(\mathbf{r},\mathbf{s},\omega)\cdot\mathbfcal{G}^{\dagger}(\mathbf{r'},\mathbf{s},\omega)
    \\
    =\frac{\hbar\omega^2}{\pi\epsilon_0 c^2}\mathbf{G}_{\textup{AH}}(\mathbf{r},\mathbf{r'},\omega),
\end{multline}
where the discrete index $n$ has been substituted by the continuous frequency $\omega$ and spatial indices $\mathbf{s}$, with the spatial coordinate already integrated over, and the subscript $\mathrm{AH}$ denotes the anti-Hermitian part defined via $\mathbf{G}_{\textup{AH}}(\mathbf{r},\mathbf{r'},\omega) \equiv [\mathbf{G}(\mathbf{r},\mathbf{r'},\omega) - \mathbf{G}^\dagger(\mathbf{r'},\mathbf{r},\omega)]/(2\mathrm{i})$. The second equality in \autoref{eq:em_tensor_mqed} is proved in Ref.~\citenum{Buhmann2012}.

This important result permits us to write the effective couplings in \autoref{eq:effcoupling} in terms of the EM Green's tensor:
\begin{subequations}\label{eq:effcoupling_mqed}
\begin{align}
\boldsymbol{\lambda}^{ee}\left(\mathbf{r},\mathbf{r'}\right) &= \frac{1}{\pi\epsilon_0c^2}\int_0^{\infty}d\omega \hspace{0.03cm}\omega \Re \left[\mathbf{G}_{\textup{AH}}(\mathbf{r},\mathbf{r'},\omega)\right], \\
\boldsymbol{\lambda}^{em}\left(\mathbf{r},\mathbf{r'}\right) &= - \frac{1}{\pi\epsilon_0c^2} \int_0^{\infty}d\omega \left\lbrace\Im\left[\mathbf{G}_{\textup{AH}}(\mathbf{r},\mathbf{r'},\omega)\right] \times\overleftarrow\nabla_{r'}\right\rbrace, \\
\boldsymbol{\lambda}^{me}\left(\mathbf{r},\mathbf{r'}\right) &= \frac{1}{\pi\epsilon_0c^2} \int_0^{\infty}d\omega \left\lbrace \overrightarrow\nabla_r\times \Im\left[\mathbf{G}_{\textup{AH}}(\mathbf{r},\mathbf{r'},\omega)\right]\right\rbrace, \\
\boldsymbol{\lambda}^{mm}\left(\mathbf{r},\mathbf{r'}\right) &= \frac{1}{\pi\epsilon_0c^2} \int_0^{\infty}d\omega \left\lbrace \overrightarrow\nabla_r\times \frac{1}{\omega} \Re\left[\mathbf{G}_{\textup{AH}}(\mathbf{r},\mathbf{r'},\omega)\right] \times\overleftarrow\nabla_{r'}\right\rbrace,
\end{align}
\end{subequations}
and, in the same fashion, the diamagnetic tensor field in \autoref{eq:Omega_tilde} as:
\begin{equation}\label{eq:Omega_mqed}
    \boldsymbol{\Omega}\left(\mathbf{r},\mathbf{r'}\right) = \frac{\hbar}{\pi\epsilon_0c^2} \int_0^{\infty}d\omega \left\lbrace \overrightarrow\nabla_r\times \Re \left[\mathbf{G}_{\textup{AH}}(\mathbf{r},\mathbf{r'},\omega)\right]\times\overleftarrow\nabla_{r'}\right\rbrace.
\end{equation}
Macroscopic QED thus provides a natural way to write the effective Hamiltonian in \autoref{eq:Heff} in terms of the EM Green's tensor, resulting in a practical approach to compute the effect of general optical environments on the matter system.

In order to perform the remaining integral over frequencies, we rewrite the integrands by using the relations:
\begin{subequations}\label{eq:AH_Green}
\begin{align}
    \Re\left[\mathbf{G}_{\textup{AH}}(\mathbf{r},\mathbf{r'},\omega)\right] &= \frac{1}{2}\Im\left[\mathbf{G}(\mathbf{r},\mathbf{r'},\omega)+\mathbf{G}^{\textup{T}}(\mathbf{r'},\mathbf{r},\omega)\right],\\
    \Im\left[\mathbf{G}_{\textup{AH}}(\mathbf{r},\mathbf{r'},\omega)\right] &= -\frac{1}{2}\Re\left[\mathbf{G}(\mathbf{r},\mathbf{r'},\omega)-\mathbf{G}^{\textup{T}}(\mathbf{r'},\mathbf{r},\omega)\right],
\end{align}
\end{subequations}
which are obtained by introducing the decomposition $\mathbf{G}(\mathbf{r},\mathbf{r'},\omega)=\Re\mathbf{G}(\mathbf{r},\mathbf{r'},\omega) + i \Im\mathbf{G}(\mathbf{r},\mathbf{r'},\omega)$ in the definition $\mathbf{G}_{\textup{AH}}(\mathbf{r},\mathbf{r'},\omega)=\left[\mathbf{G}(\mathbf{r},\mathbf{r'},\omega)-\mathbf{G}^{\dagger}(\mathbf{r'},\mathbf{r},\omega)\right]/\left(2i\right)$. 
The integrals appearing in the effective couplings are straightforwardly computed by using the identities $\Re z=\frac{1}{2}\left(z+z^{*}\right)$ and $\Im z=\frac{1}{2i}\left(z-z^{*}\right)$, and the Schwarz reflection principle $\mathbf{G}^{*}(\mathbf{r},\mathbf{r'},\omega)=\mathbf{G}(\mathbf{r},\mathbf{r'},-\omega)$, which permits to extend the integrals over the full real frequency line. The integrals can then be computed using contour integration techniques, considering the analytical properties of the Green's tensor, which is an analytic function in the upper complex half plane with the only possible pole on the real axis being at $\omega=0$. More detailed analysis (see Appendix~\ref{app:B} for details) reveals that all integrals have at most a first-order pole at the origin, such that the integrals converge and can be computed using the residue theorem. The effective couplings then simplify to:
\begin{widetext}
\begin{subequations}\label{eq:effcoupling_mqed_2}
\begin{align}
\boldsymbol{\lambda}^{ee}\left(\mathbf{r},\mathbf{r'}\right) &= \frac{1}{4\epsilon_0c^2}\left[\omega^2\mathbf{G}(\mathbf{r},\mathbf{r'},\omega)+\omega^2\mathbf{G}^{\textup{T}}(\mathbf{r'},\mathbf{r},\omega)\right]_{\omega=0}+\frac{1}{2\varepsilon_0}\boldsymbol{\delta}(\mathbf{r}-\mathbf{r'}), \label{eq:effcoupling_mqed_2_ee} \\
\boldsymbol{\lambda}^{em}\left(\mathbf{r},\mathbf{r'}\right) &= \frac{i}{4\epsilon_0c^2}\left\lbrace\frac{d}{d\omega}\left[\omega^2\mathbf{G}(\mathbf{r},\mathbf{r'},\omega)-\omega^2\mathbf{G}^{\textup{T}}(\mathbf{r'},\mathbf{r},\omega)\right]\right\rbrace_{\omega=0}\times\overleftarrow\nabla_{r'}, \\
\boldsymbol{\lambda}^{me}\left(\mathbf{r},\mathbf{r'}\right) &= -\frac{i}{4\epsilon_0c^2}\overrightarrow\nabla_r\times\left\lbrace\frac{d}{d\omega}\left[\omega^2\mathbf{G}(\mathbf{r},\mathbf{r'},\omega)-\omega^2\mathbf{G}^{\textup{T}}(\mathbf{r'},\mathbf{r},\omega)\right]\right\rbrace_{\omega=0}, \\
\boldsymbol{\lambda}^{mm}\left(\mathbf{r},\mathbf{r'}\right) &= \frac{1}{8\epsilon_0c^2}\overrightarrow\nabla_r\times\left\lbrace\frac{d^2}{d\omega^2}\left[\omega^2\mathbf{G}(\mathbf{r},\mathbf{r'},\omega)+\omega^2\mathbf{G}^{\textup{T}}(\mathbf{r'},\mathbf{r},\omega)\right]\right\rbrace_{\omega=0}\times\overleftarrow\nabla_{r'}.
\end{align}
\end{subequations}
\end{widetext}

All Green's tensors are evaluated at zero frequency, revealing the electrostatic and magnetostatic nature of the cavity-induced effective interactions between low-energy matter excitations even when considering the general beyond-dipole linear light-matter interaction in \autoref{eq:H}. These interaction kernels constitute the main result of the current paper. They demonstrate that the electrostatic and magnetostatic nature of low-energy matter-matter interactions mediated by an arbitrary ``cavity'' are independent of the level of approximation used for the light-matter coupling, as long as the full multimode nature of the EM environment is considered. Remarkably, within the effective Hamiltonian \autoref{eq:Heff}, the delta function in \autoref{eq:effcoupling_mqed_2_ee} exactly cancels out the polarization self-energy $\frac{1}{2\varepsilon_0}\sum_i\int d^3r\hspace{0.07cm}\hat{\mathbf{P}}_i^2(\mathbf{r})$ included in the renormalized bare low-energy matter Hamiltonian $\hat{\tilde{H}}_{le}$ within the Power-Zienau-Woolley picture used here, such that the effective Hamiltonian resembles the minimal-coupling form with environment-modified electrostatic and magnetostatic interactions. In \autoref{sec:free_space} below, we show explicitly that in the absence of media, the well-known free-space interactions are recovered.

Regarding the renormalization given by the non-linear diamagnetic term, the previous procedure does not work for evaluating the integral $\int_0^{\infty}d\omega \Im\left[\mathbf{G}(\mathbf{r},\mathbf{r'},\omega)+\mathbf{G}^{\textup{T}}(\mathbf{r'},\mathbf{r},\omega)\right]$, as it cannot be rewritten as a frequency integral over the full real line. Thus, the final form for the diamagnetic tensor field is
\begin{widetext}
    \begin{equation}\label{eq:Omega_mqed_2}
        \boldsymbol{\Omega}\left(\mathbf{r},\mathbf{r'}\right) = \frac{\hbar}{2\pi\epsilon_0c^2} \int_0^{\infty}d\omega \left\lbrace \overrightarrow\nabla_r \times \Im\left[\mathbf{G}(\mathbf{r},\mathbf{r'},\omega) + \mathbf{G}^{\textup{T}}(\mathbf{r'},\mathbf{r},\omega)\right] \times\overleftarrow\nabla_{r'} \right\rbrace,
    \end{equation}
\end{widetext}
and does not simplify to a term evaluated at zero frequency. Nevertheless, it remains a non-resonant term, as it is not frequency selective. This feature can be understood from the interpretation of the diamagnetic tensor field as the equal-time correlation function of the magnetic field over the EM vacuum state, $\boldsymbol{\Omega}\left(\mathbf{r},\mathbf{r'}\right)=\mathrm{Re}[\left\langle 0| \hat{\mathbf{B}}\left(\mathbf{r}\right)\otimes\hat{\mathbf{B}}\left(\mathbf{r'}\right)\right |0\rangle]$, which is not evaluated at any particular frequency. Consequently, it is necessary to consider the entire frequency spectrum. 

Further insight on the contribution of the diamagnetic renormalization can be provided by comparing its order of magnitude with the other cavity-mediated interactions between polarization and magnetization densities. This comparison is clearer and easier to interpret if we write the effective Hamiltonian in \autoref{eq:Heff} at lower order (dipole) in the multipole expansion of the densities $\hat{\mathbf{P}}_i(\mathbf{r})$, $\hat{\mathbf{M}}_i(\mathbf{r})$ and $\hat{\boldsymbol{\Theta}}_{\gamma}(\mathbf{r})$, as well as neglecting the Röntgen magnetization in $\hat{\mathbf{M}}_i(\mathbf{r})$ and considering only one matter element. In this limit, the effective Hamiltonian reads
\begin{multline}\label{eq:Heff_dipole}
    \hat{H}^{\textup{D}}_{\textup{eff}}=\hat{\tilde{H}}_{le}^{\textup{D}} - \hat{\mathbf{d}}\cdot \boldsymbol{\lambda}^{ee}\left(\mathbf{r}_i,\mathbf{r}_i\right)\cdot\hat{\mathbf{d}} - \hat{\mathbf{d}}\cdot \boldsymbol{\lambda}^{em}\left(\mathbf{r}_i,\mathbf{r}_i\right)\cdot\hat{\mathbf{m}}
    \\ - \hat{\mathbf{m}}\cdot \boldsymbol{\lambda}^{me}\left(\mathbf{r}_i,\mathbf{r}_i\right)\cdot\hat{\mathbf{d}} - \hat{\mathbf{m}}\cdot \boldsymbol{\lambda}^{mm}\left(\mathbf{r}_i,\mathbf{r}_i\right)\cdot\hat{\mathbf{m}},
\end{multline}
\noindent where the superscript $D$ stands for dipole approximation, and the electric and magnetic dipole moments are given by $\hat{\mathbf{d}}=\sum_{\gamma}q_{\gamma}\hat{\tilde{\mathbf{r}}}_{\gamma}$  and $\hat{\mathbf{m}}=\sum_{\gamma} \frac{q_{\gamma}}{2} \hat{\tilde{\mathbf{r}}}_{\gamma}\times\dot{\hat{\tilde{\mathbf{r}}}}_{\gamma}$, respectively, with $q_{\gamma}$ being the charge of the $\gamma$ constituent of the matter element and $\hat{\tilde{\mathbf{r}}}$ its position operator relative to the centre-of-mass position $\mathbf{r}_i$ of the matter element. The only change in the corrected bare matter Hamiltonian, $\hat{\tilde{H}}_{le}^{\textup{D}}$, involves the diamagnetic renormalization contribution, which in this approximation~\cite{Buhmann2012BI} takes the form:
\begin{widetext}
     \begin{equation}\label{eq:Hren_dipole}
        \hat{H}_{ren}^{\textup{D}} = \sum_{\gamma}\frac{q_{\gamma}^2}{16\pi^2\epsilon_0} \textup{Tr}\left\lbrace\hat{\tilde{\mathbf{r}}}_{\gamma}\times\left[\frac{\lambda_{\gamma}}{c}\hspace{0.05cm} \int_0^{\infty}d\omega \left\lbrace \overrightarrow\nabla_r\times \Im\left[\mathbf{G}(\mathbf{r},\mathbf{r'},\omega)+\mathbf{G}^{\textup{T}}(\mathbf{r'},\mathbf{r},\omega)\right] \times\overleftarrow\nabla_{r'} \right\rbrace\right]_{\mathbf{r}=\mathbf{r}_i,\mathbf{r'}=\mathbf{r}_i}\times\hat{\tilde{\mathbf{r}}}_{\gamma}\right\rbrace,
    \end{equation}
\end{widetext}
where we have introduced the Compton wavelength, defined by $\lambda_{\gamma}=2\pi\hbar/(m_{\gamma}c)$ with $m_{\gamma}$ the mass of the $\gamma$ constituent of the matter element. By identifying $\hat{H}_{ee}^{\textup{D}}=\hat{\mathbf{d}}\cdot \boldsymbol{\lambda}^{ee}\left(\mathbf{r}_i,\mathbf{r}_i\right)\cdot\hat{\mathbf{d}}$, $\hat{H}_{em}^{\textup{D}}=\hat{\mathbf{d}}\cdot \boldsymbol{\lambda}^{em}\left(\mathbf{r}_i,\mathbf{r}_i\right)\cdot\hat{\mathbf{m}}$, $\hat{H}_{me}^{\textup{D}}=\hat{\mathbf{m}}\cdot \boldsymbol{\lambda}^{me}\left(\mathbf{r}_i,\mathbf{r}_i\right)\cdot\hat{\mathbf{d}}$, and $\hat{H}_{mm}^{\textup{D}}=\hat{\mathbf{m}}\cdot \boldsymbol{\lambda}^{mm}\left(\mathbf{r}_i,\mathbf{r}_i\right)\cdot\hat{\mathbf{m}}$, we can write
\begin{widetext}
\begin{subequations}\label{eq:effcoupling_dipole}
\begin{align}
\hat{H}_{ee}^{\textup{D}} &= \sum_{\gamma\gamma'}\frac{q_{\gamma}q_{\gamma'}}{4\epsilon_0}\hat{\tilde{\mathbf{r}}}_{\gamma}\cdot\left[\frac{\omega^2}{c^2}\mathbf{G}(\mathbf{r}_i,\mathbf{r}_i,\omega)+\frac{\omega^2}{c^2}\mathbf{G}^{\textup{T}}(\mathbf{r}_i,\mathbf{r}_i,\omega)\right]_{\omega=0}\cdot\hat{\tilde{\mathbf{r}}}_{\gamma'} + \frac{1}{2\varepsilon_0}\sum_{\gamma}q_{\gamma}^2\hspace{0.03cm}\hat{\tilde{\mathbf{r}}}^2\delta(\mathbf{0}),\\
\hat{H}_{em}^{\textup{D}} &= \sum_{\gamma\gamma'}i\frac{q_{\gamma}q_{\gamma'}}{4\epsilon_0}\hat{\tilde{\mathbf{r}}}_{\gamma}\cdot\left[\left\lbrace\frac{d}{d\omega}\left[\frac{\omega^2}{c^2}\mathbf{G}(\mathbf{r}_i,\mathbf{r'},\omega)-\frac{\omega^2}{c^2}\mathbf{G}^{\textup{T}}(\mathbf{r'},\mathbf{r}_i,\omega)\right]\right\rbrace_{\omega=0}\times\overleftarrow\nabla_{r'}\right]_{\mathbf{r'}=\mathbf{r}_i}\cdot\frac{1}{2}\hat{\tilde{\mathbf{r}}}_{\gamma'}\times\dot{\hat{\tilde{\mathbf{r}}}}_{\gamma'},\\
\hat{H}_{me}^{\textup{D}} &= -\sum_{\gamma\gamma'}i\frac{q_{\gamma}q_{\gamma'}}{4\epsilon_0}\frac{1}{2}\hat{\tilde{\mathbf{r}}}_{\gamma}\times\dot{\hat{\tilde{\mathbf{r}}}}_{\gamma}\cdot\left[\overrightarrow\nabla_r\times\left\lbrace\frac{d}{d\omega}\left[\frac{\omega^2}{c^2}\mathbf{G}(\mathbf{r},\mathbf{r}_i,\omega)-\frac{\omega^2}{c^2}\mathbf{G}^{\textup{T}}(\mathbf{r}_i,\mathbf{r},\omega)\right]\right\rbrace_{\omega=0}\right]_{\mathbf{r}=\mathbf{r}_i}\cdot\hat{\tilde{\mathbf{r}}}_{\gamma'},\\
\hat{H}_{mm}^{\textup{D}} &= \sum_{\gamma\gamma'}\frac{q_{\gamma}q_{\gamma'}}{8\epsilon_0}\frac{1}{2}\hat{\tilde{\mathbf{r}}}_{\gamma}\times\dot{\hat{\tilde{\mathbf{r}}}}_{\gamma}\cdot\left[\overrightarrow\nabla_r\times\left\lbrace\frac{d^2}{d\omega^2}\left[\frac{\omega^2}{c^2}\mathbf{G}(\mathbf{r},\mathbf{r'},\omega)+\frac{\omega^2}{c^2}\mathbf{G}^{\textup{T}}(\mathbf{r'},\mathbf{r},\omega)\right]\right\rbrace_{\omega=0}\times\overleftarrow\nabla_{r'}\right]_{\mathbf{r}=\mathbf{r}_i,\mathbf{r'}=\mathbf{r}_i}\cdot\frac{1}{2}\hat{\tilde{\mathbf{r}}}_{\gamma'}\times\dot{\hat{\tilde{\mathbf{r}}}}_{\gamma'},
\end{align}
\end{subequations}
\end{widetext}
and establish a comparison between these terms and the diamagnetic renormalization. By following the argument in Ref.~\citenum{Barcellona2018}, note that the Green's tensor $\mathbf{G}(\mathbf{r},\mathbf{r'},\omega)$ scales as $r^{-1}$ with $r$ being the typical distance between the charges and the surface of the cavity. Therefore, the function $(\omega^2/c^2)\mathbf{G}(\mathbf{r},\mathbf{r'},\omega)$ appearing in \autoref{eq:effcoupling_dipole} has an order of magnitude of $r^{-3}$, while the function $(\lambda_{\gamma}/c)\mathbf{G}(\mathbf{r},\mathbf{r'},\omega)$ defining the diamagnetic term in \autoref{eq:Hren_dipole} scales as $\lambda_{\gamma}r^{-2}$. Thus, the ratio between $\hat{H}_{ren}^{\textup{D}}$ and the terms in \autoref{eq:effcoupling_dipole} has an order of magnitude of $\lambda_{\gamma}/r$. As the Compton wavelength for electrons fulfills $\lambda_{\gamma}\approx 10^{-12}\hspace{0.03cm}\textup{m}$ and the macroscopic description of the cavity that underpins macroscopic QED ceases to be valid for distances smaller than $10^{-9}\hspace{0.03cm}\textup{m}$, we can safely neglect the diamagnetic renormalization in our systems of interest.

\section{Limit of reciprocal media}

It is instructive to investigate the predictions of our theoretical framework in the limit of reciprocal media, while still accounting for the possibility of non-local response. Note that a spatially local description of typical conducting, semiconducting and superconducting materials, where charges move almost freely, may not be correct, especially when the matter resides close to these materials~\cite{Ciraci2013}. Instead, reciprocal media is the most common situation encountered in cavity QED platforms~\cite{Haroche2007}. Physically, reciprocity describes a symmetry with respect to an exchange of positions and orientations of sources and fields, i.e., the reversibility of optical paths. 

In reciprocal media, one finds that the Green tensor satisfies the Onsager reciprocity condition $\mathbf{G}^{\textup{T}}(\mathbf{r'},\mathbf{r},\omega)=\mathbf{G}(\mathbf{r},\mathbf{r'},\omega)$. Using this relation, the effective couplings take the form:
\begin{subequations}\label{eq:effcoupling_mqed_reciprocal}
\begin{align}
\boldsymbol{\lambda}^{ee}\left(\mathbf{r},\mathbf{r'}\right) &= \frac{1}{2\epsilon_0c^2}\left[\omega^2\mathbf{G}(\mathbf{r},\mathbf{r'},\omega)\right]_{\omega=0}, \label{eq:pol_pol_reciprocal} \\
\boldsymbol{\lambda}^{em}\left(\mathbf{r},\mathbf{r'}\right) &= 0, \\
\boldsymbol{\lambda}^{me}\left(\mathbf{r},\mathbf{r'}\right) &= 0, \\
\boldsymbol{\lambda}^{mm}\left(\mathbf{r},\mathbf{r'}\right) &= \frac{1}{4\epsilon_0c^2}\overrightarrow\nabla_r\times\left\lbrace\frac{d^2}{d\omega^2}\left[\omega^2\mathbf{G}(\mathbf{r},\mathbf{r'},\omega)\right]\right\rbrace_{\omega=0}\times\overleftarrow\nabla_{r'}. \label{eq:mag_mag_reciprocal}
\end{align}
\end{subequations}

As the cross-couplings vanish, the most important implication of reciprocity is that there are no cross electric-magnetic transitions, eliminating any chiral effect in the effective Hamiltonian.
We note that this is the case even if there are magnetoelectric media in the system, as long as they are reciprocal.

\section{Limit of free space}\label{sec:free_space}

As an illustration of \autoref{eq:effcoupling_mqed_reciprocal}, we study the free-space case, where the EM Green's tensor is given by: 
\begin{multline}\label{eq:G0}
    \mathbf{G}_0(\mathbf{r}, \mathbf{r'}, \omega)=\left[\mathbf{I}+\frac{1}{k^2}\overrightarrow\nabla_r \otimes \overrightarrow\nabla_r\right]\frac{e^{ikR}}{4\pi R} = -\frac{\mathbf{n}_{R}\otimes\mathbf{n}_{R}}{k^2}\delta(\mathbf{R})
    \\
    +\frac{e^{ikR}}{4\pi R}\left[\mathbf{I}-\mathbf{n}_{R}\otimes\mathbf{n}_{R}+\frac{\left(ikR-1\right)}{k^2R^2}\left(\mathbf{I}-3\mathbf{n}_{R}\otimes\mathbf{n}_{R}\right)\right],
\end{multline}
where $\mathbf{R}=\mathbf{r}-\mathbf{r'}$, $R=\abs{\mathbf{R}}$, $\mathbf{n}_R=\mathbf{R}/R$, and $k=\omega/c$. As free space is reciprocal, the effective electric-magnetic and magnetic-electric cross-couplings vanish. The first term containing a delta function arises because of the second derivative on the inverse distance $1/R$~\cite{Franklin2010}. By inserting \autoref{eq:G0} into \autoref{eq:pol_pol_reciprocal}, we find that the free-space electric-electric effective coupling reads
\begin{equation}\label{eq:lambda_0}
    \boldsymbol{\lambda}_0^{ee}\left(\mathbf{r},\mathbf{r'}\right)=\frac{1}{2\varepsilon_0}\left[-\mathbf{n}_{R}\otimes\mathbf{n}_{R}\delta(\mathbf{R}) + \frac{3\mathbf{n}_{R}\otimes\mathbf{n}_{R}-\mathbf{I}}{4\pi R^3}\right]  + \frac{1}{2\varepsilon_0}\boldsymbol{\delta}(\mathbf{R}).
\end{equation}
We note that that expressions containing $\delta(\mathbf{R})$ are to be understood as functionals acting on another function within an integral, in this case the polarization density. As long as the polarization densities are smooth functions, the equivalence $\mathbf{n}_{R}\otimes\mathbf{n}_{R}\delta(\mathbf{R}) = \frac{1}{3}\boldsymbol{\delta}(\mathbf{R})$ holds~\cite{Franklin2010}, showing the emergence of the well-known $1/3$ factor in the delta-term of the kernel of the free-space electrostatic interaction energy. The overall factor $1/2$ accounts for the fact that each pair of polarizations appears twice in the sum over $i$ and $j$. As mentioned above, the final term of \autoref{eq:lambda_0} cancels the polarization self-energy included in the bare matter Hamiltonian, showing the recovery of the standard Coulomb interaction as directly obtained in the minimal coupling scheme. 

The evaluation of the free-space magnetic-magnetic effective coupling is more involved. First, we compute:
\begin{equation}
    \left\lbrace\frac{d^2}{d\omega^2}\left[\omega^2\hspace{0.05cm} \mathbf{G}_0\left(\mathbf{r},\mathbf{r'},\omega\right)\right]\right\rbrace_{\omega=0}=\frac{1}{4\pi R}\left(\mathbf{I}+\mathbf{n}_{R}\otimes\mathbf{n}_{R}\right).
\end{equation}
Then, by using the definition of the two-side curl, it can be shown that the free-space magnetic-magnetic effective coupling reduces to the well-known kernel of the free-space magnetostatic interaction energy:
\begin{equation}
    \boldsymbol{\lambda}_0^{mm}\left(\mathbf{r},\mathbf{r'}\right)=\frac{\mu_0}{2}\left[\left(\mathbf{I}-\mathbf{n}_{R}\otimes\mathbf{n}_{R}\right)\delta(\mathbf{R}) + \frac{3\mathbf{n}_{R}\otimes\mathbf{n}_{R}-\mathbf{I}}{4\pi R^3}\right].
\end{equation}
For smooth magnetization densities, the equivalence given above leads to the standard factor $2/3$ in the delta-function part of the kernel.

This demonstrates how our formalism recovers thefree-space limit, in which no cavity is considered.

\section{Discussion and conclusions}

We have studied the direct interaction between low-energy matter excitations and an arbitrary cavity EM field, under general conditions of both cavity media and light-matter coupling. We have considered non-local and non-reciprocal media, and the full polarization and magnetization densities of matter, as well as the diamagnetic contribution. Our analysis has confirmed the zero-frequency (i.e., electrostatic-magnetostatic) nature of the four effective induced couplings between polarization and magnetization densities [\autoref{eq:effcoupling_mqed_2}], and revealed the contribution coming from the diamagnetic term as a non-frequency-selective energy shift [\autoref{eq:Omega_mqed_2}]. Note that this procedure can be further generalized to consider bosonic mediator modes as ref.~\citenum{Pantazopoulos2024}. 

We highlight the non-resonant behavior of the obtained effective Hamiltonian for the low-energy matter degrees of freedom, independently of the level of approximation in the matter coupling. This result demonstrates the importance of considering the full multimode nature of the EM field in the off-resonant regime, especially indicating that results obtained in a single-mode approximation have to be considered with care. Moreover, the presented generalization of the formalism to extended matter excitations provides a recipe for studying the field-induced interactions in nanophotonics platforms with strong field gradients, where the electric-dipole approximation breaks down. Finally, the electrostatic-magnetostatic nature of the effective induced interactions imposes a clear limit on the possibilities of cavity engineering on matter systems where the low-energy assumption applies.

\begin{acknowledgments}
This work was funded by the Spanish Ministry for Science and Innovation-Agencia Estatal de Investigación (AEI) through FPI contract No.~PRE2022-101819 as well as grants PID2021-125894NB-I00, EUR2023-143478, and CEX2018-000805-M (through the María de Maeztu program for Units of Excellence in R\&D). We also acknowledge financial support from the Proyecto Sinérgico CAM 2020 Y2020/TCS-6545 (NanoQuCo-CM) of the Community of Madrid, and from the European Union’s Horizon Europe Research and Innovation Programme through agreements 101070700 (MIRAQLS) and 101098813 (SCOLED).
\end{acknowledgments}

\bibliography{references}



\appendix

\section{\label{app:A} Derivation of the effective Hamiltonian}

The starting point is the formal definition of the effective Hamiltonian in \autoref{eq:Heff}. The first step is to compute the multi-coherent-state expectation value $\bra{\alpha}\exp\left(-\beta\hat{H}\right)\ket{\alpha}$ in the thermodynamic limit~\cite{Roman-Roche2022,Coleman2015}:
\begin{multline}\label{eq:Heff_a1_v1}
    \exp\left(-\beta\hat{H}_{\textup{eff}}\right)=\prod_n\left(\int\frac{d^2\alpha_n}{\pi}\right)\exp\left\lbrace -\beta\left[\hat{H}_{le}+\sum_n\hbar\omega_n\abs{\alpha_n}^2 \right.\right.
    \\
    \left. \left. -\sum_n\int d^3r\sum_i\hat{\mathbf{P}}_i(\mathbf{r})\cdot\mathbf{E}_n(\mathbf{r})\alpha_n -\sum_n\int d^3r\sum_i\hat{\mathbf{P}}_i(\mathbf{r})\cdot\mathbf{E}_n^{*}(\mathbf{r})\alpha_n^{*}  \right. \right.
    \\
    \left. \left. -\sum_n\int d^3r\sum_i\hat{\mathbf{M}}_i(\mathbf{r})\cdot\mathbf{B}_n(\mathbf{r})\alpha_n -\sum_n\int d^3r\sum_i\hat{\mathbf{M}}_i(\mathbf{r})\cdot\mathbf{B}_n^{*}(\mathbf{r})\alpha_n^{*} \right. \right.  \\
    \left. \left. + \sum_{nm}\int d^3r\int d^3r'\sum_i\sum_{\gamma\in i} \right.\right.
    \\
    \left.\left. \left(\textup{Tr}\left[\hat{\boldsymbol{\Theta}}_{\gamma}\left(\mathbf{r}\right)\times\mathbf{B}_n(\mathbf{r})\otimes\mathbf{B}_m(\mathbf{r'})\times\hat{\boldsymbol{\Theta}}_{\gamma}\left(\mathbf{r'}\right)\right]\alpha_n\alpha_m \right.\right.\right.
    \\
    \left.\left.\left. + \textup{Tr}\left[\hat{\boldsymbol{\Theta}}_{\gamma}\left(\mathbf{r}\right)\times\mathbf{B}_n(\mathbf{r})\otimes\mathbf{B}_m^{*}(\mathbf{r'})\times\hat{\boldsymbol{\Theta}}_{\gamma}\left(\mathbf{r'}\right)\right]\delta_{nm} \right.\right.\right.
    \\
    \left.\left.\left. + \textup{Tr}\left[\hat{\boldsymbol{\Theta}}_{\gamma}\left(\mathbf{r}\right)\times\mathbf{B}_n(\mathbf{r})\otimes\mathbf{B}_m^{*}(\mathbf{r'})\times\hat{\boldsymbol{\Theta}}_{\gamma}\left(\mathbf{r'}\right)\right]\alpha_n\alpha_m^{*} \right.\right.\right.
    \\
    \left.\left.\left. + \textup{Tr}\left[\hat{\boldsymbol{\Theta}}_{\gamma}\left(\mathbf{r}\right)\times\mathbf{B}_n^{*}(\mathbf{r})\otimes\mathbf{B}_m(\mathbf{r'})\times\hat{\boldsymbol{\Theta}}_{\gamma}\left(\mathbf{r'}\right)\right]\alpha_n^{*}\alpha_m \right.\right.\right.
    \\
    \left. + \textup{Tr}\left[\hat{\boldsymbol{\Theta}}_{\gamma}\left(\mathbf{r}\right)\times\mathbf{B}_n^{*}(\mathbf{r})\otimes\mathbf{B}_m^{*}(\mathbf{r'})\times\hat{\boldsymbol{\Theta}}_{\gamma}\left(\mathbf{r'}\right)\right]\alpha_n^{*}\alpha_m^{*} \right)\bigg ] \bigg\rbrace.
\end{multline}
Here, we have used the definition of the electric and magnetic operators: $\hat{\mathbf{E}}\left(\mathbf{r}\right)=\sum_n\left[\mathbf{E}_n\left(\mathbf{r} \right)\hat{a}_n + \mathbf{E}^{*}_n\left(\mathbf{r} \right)\hat{a}^{\dagger}_n\right]$ and $\hat{\mathbf{B}}\left(\mathbf{r}\right)=\sum_n\left[\mathbf{B}_n\left(\mathbf{r} \right)\hat{a}_n + \mathbf{B}^{*}_n\left(\mathbf{r} \right)\hat{a}^{\dagger}_n\right]$, together with the definition of the annihilation operator: $\hat{a}_n\ket{\alpha_n}=\alpha_n\ket{\alpha_n}$. In the derivation of the contribution given by the diamagnetic term, we have utilized the fact that all the operators must be written in normal order when evaluating coherent-state expectation values. To do so, we use $\hat{a}_n\hat{a}_m^{\dagger}=\delta_{nm}+\hat{a}_m^{\dagger}\hat{a}_n$. 

The next step is to evaluate the multi-dimensional Gaussian integral appearing in \autoref{eq:Heff_a1_v1}. For this purpose, we use the general formula $\int_{\mathbb{R}^{m}}\exp(-\frac{1}{2}\mathbf{x}^T\cdot\mathbf{A}\cdot\mathbf{x}+\mathbf{b}^T\cdot\mathbf{x}+c)\hspace{0.05cm}d^m x=\sqrt{\det(2\pi\mathbf{A}^{-1})}\exp(\frac{1}{2}\mathbf{b}^T\cdot\mathbf{A}^{-1}\cdot\mathbf{b}+c)$. This formula cannot be directly applied to \autoref{eq:Heff_a1_v1}; we must perform a change of variables to the real and imaginary parts of the complex eigenvalues $\alpha_n$: $\alpha_n'= \Re \alpha_n$ and $\alpha_n'' = \Im \alpha_n$. Thus, \autoref{eq:Heff_a1_v1} can be written as:
\begin{widetext}
\begin{multline}
    \exp\left(-\beta\hat{H}_{\textup{eff}}\right)=\prod_n\left(\int\frac{d\alpha_n'd\alpha_n''}{\pi}\right) \\
    \times \exp\bigg\lbrace -\beta\bigg[\hat{H}_{le} + \sum_{n}\int d^3r\int d^3r'\sum_{i,\gamma\in i}\textup{Tr}\left[\hat{\boldsymbol{\Theta}}_{\gamma}\left(\mathbf{r}\right)\times\mathbf{B}_n(\mathbf{r})\otimes\mathbf{B}_n^{*}(\mathbf{r'})\times\hat{\boldsymbol{\Theta}}_{\gamma}\left(\mathbf{r'}\right)\right]+\sum_n \hbar \omega_n\left(\alpha_{n}^{\prime2}+\alpha_{n}^{\prime\prime2}\right) 
    \\
    \left.\left. - 2\sum_n \sum_i \int d^3r \left\lbrace \hat{\mathbf{P}}_i\left(\mathbf{r}\right)\cdot \Re\left[\mathbf{E}_n\left(\mathbf{r}\right)\right] + \hat{\mathbf{M}}_i\left(\mathbf{r}\right)\cdot \Re\left[\mathbf{B}_n\left(\mathbf{r}\right)\right] \right\rbrace\alpha_n^{\prime} + 2\sum_n\sum_i \int d^3r \left\lbrace \hat{\mathbf{P}}_i\left(\mathbf{r}\right)\cdot \Im\left[\mathbf{E}_n\left(\mathbf{r}\right)\right] + \hat{\mathbf{M}}_i\left(\mathbf{r}\right)\cdot \Im\left[\mathbf{B}_n\left(\mathbf{r}\right) \right]\right\rbrace\alpha_n^{\prime\prime} \right.\right.
    \\
    \left.\left. + 2\sum_{nm}\int d^3r\int d^3r'\sum_{i,\gamma\in i}\left(\textup{Tr}\left\lbrace\hat{\boldsymbol{\Theta}}_{\gamma}\left(\mathbf{r}\right)\times\Re\left[\mathbf{B}_n(\mathbf{r})\otimes\mathbf{B}_m(\mathbf{r'})+\mathbf{B}_n(\mathbf{r})\otimes\mathbf{B}_m^{*}(\mathbf{r'})\right]\times\hat{\boldsymbol{\Theta}}_{\gamma}\left(\mathbf{r'}\right)\right\rbrace\alpha_n'\alpha_m' \right.\right.\right.
    \\
    \left.\left.\left. + \textup{Tr}\left\lbrace\hat{\boldsymbol{\Theta}}_{\gamma}\left(\mathbf{r}\right)\times\Re\left[-\mathbf{B}_n(\mathbf{r})\otimes\mathbf{B}_m(\mathbf{r'})+\mathbf{B}_n(\mathbf{r})\otimes\mathbf{B}_m^{*}(\mathbf{r'})\right]\times\hat{\boldsymbol{\Theta}}_{\gamma}\left(\mathbf{r'}\right)\right\rbrace\alpha_n''\alpha_m''
    \right.\right.\right.
    \\
    \left.\left.\left.- \textup{Tr}\left\lbrace\hat{\boldsymbol{\Theta}}_{\gamma}\left(\mathbf{r}\right)\times\Im\left[\mathbf{B}_n(\mathbf{r})\otimes\mathbf{B}_m(\mathbf{r'})-\mathbf{B}_n(\mathbf{r})\otimes\mathbf{B}_m^{*}(\mathbf{r'})\right]\times\hat{\boldsymbol{\Theta}}_{\gamma}\left(\mathbf{r'}\right)\right\rbrace\alpha_n'\alpha_m'' \right.\right.\right.
    \\
    \left. - \textup{Tr}\left\lbrace\hat{\boldsymbol{\Theta}}_{\gamma}\left(\mathbf{r}\right)\times\Im\left[\mathbf{B}_n(\mathbf{r})\otimes\mathbf{B}_m(\mathbf{r'})+\mathbf{B}_n(\mathbf{r})\otimes\mathbf{B}_m^{*}(\mathbf{r'})\right]\times\hat{\boldsymbol{\Theta}}_{\gamma}\left(\mathbf{r'}\right)\right\rbrace\alpha_n''\alpha_m'
    \right)\bigg]\bigg\rbrace.
\end{multline}
\end{widetext}
This allows to identify
\begin{subequations}
\begin{align}
    \mathbf{x}^T &= \left(\alpha_1^{\prime},\alpha_1^{\prime\prime},\alpha_2^{\prime},\alpha_2^{\prime\prime},\ldots \right), \\
    \mathbf{b}^T &= 2\beta\hspace{0.05cm}\left(\hat{L}_{1}^{Re},-\hat{L}_{1}^{Im},\hat{L}_{2}^{Re},-\hat{L}_{2}^{Im},\ldots \right), \\
    \begin{split}
        c &= -\beta\bigg[\hat{H}_{le} + \sum_{n}\int d^3r\int d^3r'\sum_{i,\gamma\in i} 
        \\
        &\times\textup{Tr}\left[\hat{\boldsymbol{\Theta}}_{\gamma}\left(\mathbf{r}\right)\times\mathbf{B}_n(\mathbf{r})\otimes\mathbf{B}_n^{*}(\mathbf{r'})\times\hat{\boldsymbol{\Theta}}_{\gamma}\left(\mathbf{r'}\right)\right]\bigg],\label{eq:c}
    \end{split}
\end{align}
\end{subequations}
with $\hat{L}_n^{Re}\equiv\sum_i\int d^3r\left\lbrace\hat{\mathbf{P}}_i\left(\mathbf{r}\right)\cdot \Re\left[\mathbf{E}_n\left(\mathbf{r}\right)\right] + \hat{\mathbf{M}}_i\left(\mathbf{r}\right)\cdot \Re\left[\mathbf{B}_n\left(\mathbf{r}\right)\right]\right\rbrace$ and $\hat{L}_n^{Im}\equiv\sum_i\int d^3r\left\lbrace\hat{\mathbf{P}}_i\left(\mathbf{r}\right)\cdot \Im\left[\mathbf{E}_n\left(\mathbf{r}\right)\right] + \hat{\mathbf{M}}_i\left(\mathbf{r}\right)\cdot \Im\left[\mathbf{B}_n\left(\mathbf{r}\right)\right]\right\rbrace$. The remaining terms, namely the bare EM energy and the diamagnetic contribution, are included in the matrix $\mathbf{A}$. Due to the diamagnetic contribution, the application of the general formula for multi-dimensional Gaussian integrals yields terms of fourth order in the EM field. We derive our effective Hamiltonian neglecting these high-order contributions, such that $\mathbf{A}=2\beta\hspace{0.05cm}\textup{diag}\left(\hbar\omega_1, \hbar\omega_1, \hbar\omega_2, \hbar\omega_2, \ldots \right)$. Thus, the effective Hamiltonian, up to second order in the EM field amplitude, reads as:
\begin{equation}
    \hat{H}_{\textup{eff}}=\hat{\tilde{H}}_{le} - \sum_n\left[\frac{\left(\hat{L}_n^{Re}\right)^2+\left(\hat{L}_n^{Im}\right)^2}{\hbar\omega_n}\right],
\end{equation}
where we have defined the corrected low-energy matter Hamiltonian as $\hat{\tilde{H}}_{le}=\hat{H}_{le}+\hat{H}_{ren}$, with:
\begin{equation}\label{eq:Hren_a1}
    \hat{H}_{ren} = \int d^3r\int d^3r'\sum_{i,\gamma\in i}\textup{Tr}\left[\hat{\boldsymbol{\Theta}}_{\gamma}\left(\mathbf{r}\right)\times\boldsymbol{\Omega}(\mathbf{r},\mathbf{r'})\times\hat{\boldsymbol{\Theta}}_{\gamma}\left(\mathbf{r'}\right)\right],
\end{equation}
The diamagnetic field tensor is:
\begin{equation}
    \boldsymbol{\Omega}(\mathbf{r},\mathbf{r'})=\sum_n\mathrm{Re}\left[\mathbf{B}_n(\mathbf{r})\otimes\mathbf{B}_n^{*}(\mathbf{r'})\right].
\end{equation}
Notice that the imaginary part of the tensor $\mathbf{B}_n(\mathbf{r})\otimes\mathbf{B}_n^{*}(\mathbf{r'})$ does not appear. It can be shown that $\mathrm{Im}\left[\mathbf{B}_n(\mathbf{r})\otimes\mathbf{B}_n^{*}(\mathbf{r'})\right]$ does not contribute in \autoref{eq:c}:
\begin{multline}
    \int d^3r\int d^3r'\textup{Tr}\left\lbrace\hat{\boldsymbol{\Theta}}_{\gamma}\left(\mathbf{r}\right)\times\mathrm{Im}\left[\mathbf{B}_n(\mathbf{r})\otimes\mathbf{B}_n^{*}(\mathbf{r'})\right]\times\hat{\boldsymbol{\Theta}}_{\gamma}\left(\mathbf{r'}\right)\right\rbrace
    \\
    = \frac{1}{2i}\int d^3r\int d^3r'\textup{Tr}\left\lbrace\hat{\boldsymbol{\Theta}}_{\gamma}\left(\mathbf{r}\right)\times\left[\mathbf{B}_n(\mathbf{r})\otimes\mathbf{B}_n^{*}(\mathbf{r'}) \right.\right.
    \\
    \left. \left. - \mathbf{B}_n^{*}(\mathbf{r})\otimes\mathbf{B}_n(\mathbf{r'})\right]\times\hat{\boldsymbol{\Theta}}_{\gamma}\left(\mathbf{r'}\right)\right\rbrace
    \\
    \propto \int d^3r\int d^3r'\sum_{\alpha\beta\delta\mu\nu}\left[\epsilon^{\alpha\beta\delta}\epsilon^{\alpha\mu\nu}\hat{\Theta}_{\gamma}^{\beta}(\mathbf{r})\hat{\Theta}_{\gamma}^{\mu}(\mathbf{r'})B_n^{\delta}(\mathbf{r})B_n^{\nu \hspace{0.01cm}*}(\mathbf{r'})\right.
    \\
    \left. - \epsilon^{\alpha\beta\delta}\epsilon^{\alpha\mu\nu}\hat{\Theta}_{\gamma}^{\beta}(\mathbf{r})\hat{\Theta}_{\gamma}^{\mu}(\mathbf{r'})B_n^{\delta\hspace{0.01cm}*}(\mathbf{r})B_n^{\nu }(\mathbf{r'})\right]
    \\
    = \int d^3r\int d^3r'\sum_{\alpha\beta\delta\mu\nu}\left[\epsilon^{\alpha\beta\delta}\epsilon^{\alpha\mu\nu}\hat{\Theta}_{\gamma}^{\beta}(\mathbf{r})\hat{\Theta}_{\gamma}^{\mu}(\mathbf{r'})B_n^{\delta}(\mathbf{r})B_n^{\nu \hspace{0.01cm}*}(\mathbf{r'})\right.
    \\
    \left. - \epsilon^{\alpha\mu\nu}\epsilon^{\alpha\beta\delta}\hat{\Theta}_{\gamma}^{\mu}(\mathbf{r'})\hat{\Theta}_{\gamma}^{\beta}(\mathbf{r})B_n^{\nu }(\mathbf{r'})B_n^{\delta\hspace{0.01cm}*}(\mathbf{r})\right],
\end{multline}
where we have used the commutativity of the density operator $\hat{\boldsymbol{\Theta}}(\mathbf{r})$. Then, by interchanging $\mathbf{r}\leftrightarrow\mathbf{r'}$, $\beta\leftrightarrow\mu$, $\delta\leftrightarrow\nu$, we find
\begin{multline}
    \int d^3r\int d^3r'\textup{Tr}\left\lbrace\hat{\boldsymbol{\Theta}}_{\gamma}\left(\mathbf{r}\right)\times\mathrm{Im}\left[\mathbf{B}_n(\mathbf{r})\otimes\mathbf{B}_n^{*}(\mathbf{r'})\right]\times\hat{\boldsymbol{\Theta}}_{\gamma}\left(\mathbf{r'}\right)\right\rbrace
    \\
    \propto \int d^3r\int d^3r'\sum_{\alpha\beta\delta\mu\nu}\left[\epsilon^{\alpha\beta\delta}\epsilon^{\alpha\mu\nu}\hat{\Theta}_{\gamma}^{\beta}(\mathbf{r})\hat{\Theta}_{\gamma}^{\mu}(\mathbf{r'})B_n^{\delta}(\mathbf{r})B_n^{\nu \hspace{0.01cm}*}(\mathbf{r'})\right.
    \\
    \left. - \epsilon^{\alpha\beta\delta}\epsilon^{\alpha\mu\nu}\hat{\Theta}_{\gamma}^{\beta}(\mathbf{r})\hat{\Theta}_{\gamma}^{\mu}(\mathbf{r'})B_n^{\delta }(\mathbf{r})B_n^{\nu\hspace{0.01cm}*}(\mathbf{r'})\right]=0.
\end{multline}
 By using the identity $\left(\sum_i a_i \Re b_i\right)^2+\left(\sum_i a_i \Im b_i\right)^2=\sum_{ij}a_i a_j\left(\Re b_i \Re b_j+ \Im b_i \Im b_j\right)=\sum_{ij} a_i a_j \Re(b_i b_j^{*})$, we write the effective Hamiltonian in the way of \autoref{eq:Heff}:
\begin{multline}
    \hat{H}_{\textup{eff}}=\hat{\tilde{H}}_{le} - \sum_{i,j}\int d^3r\int d^3r^{\prime}\left[\hat{\mathbf{P}}_i\left(\mathbf{r}\right)\cdot \boldsymbol{\lambda}^{ee}\left(\mathbf{r},\mathbf{r'}\right)\cdot\hat{\mathbf{P}}_j\left(\mathbf{r'}\right) \right.
    \\
    \left. + \hat{\mathbf{P}}_i\left(\mathbf{r}\right)\cdot \boldsymbol{\lambda}^{em}\left(\mathbf{r},\mathbf{r'}\right)\cdot\hat{\mathbf{M}}_j\left(\mathbf{r'}\right) \right.
    \\
    \left. + \hat{\mathbf{M}}_i\left(\mathbf{r}\right)\cdot \boldsymbol{\lambda}^{me}\left(\mathbf{r},\mathbf{r'}\right)\cdot\hat{\mathbf{P}}_j\left(\mathbf{r'}\right) \right.
    \\
    \left.+ \hat{\mathbf{M}}_i\left(\mathbf{r}\right)\cdot \boldsymbol{\lambda}^{mm}\left(\mathbf{r},\mathbf{r'}\right)\cdot\hat{\mathbf{M}}_j\left(\mathbf{r'}\right)\right],
\end{multline}
with the effective couplings defined by:
\begin{subequations}
\begin{align}
\boldsymbol{\lambda}^{ee}\left(\mathbf{r},\mathbf{r'}\right) &= \sum_n \Re \left[\frac{\mathbf{E}_n\left( \mathbf{r} \right)\otimes \mathbf{E}^{*}_n\left(\mathbf{r'} \right)}{\hbar \omega_n} \right], \\
\boldsymbol{\lambda}^{em}\left(\mathbf{r},\mathbf{r'}\right) &= \sum_n \Re \left[\frac{\mathbf{E}_n\left( \mathbf{r} \right)\otimes \mathbf{B}^{*}_n\left(\mathbf{r'}\right)}{\hbar \omega_n} \right], \\
\boldsymbol{\lambda}^{me}\left(\mathbf{r},\mathbf{r'}\right) &= \sum_n \Re \left[\frac{\mathbf{B}_n\left( \mathbf{r} \right)\otimes \mathbf{E}^{*}_n\left(\mathbf{r'} \right)}{\hbar \omega_n} \right], \\
\boldsymbol{\lambda}^{mm}\left(\mathbf{r},\mathbf{r'}\right) &= \sum_n\Re \left[\frac{\mathbf{B}_n\left( \mathbf{r} \right)\otimes\mathbf{B}^{*}_n\left(\mathbf{r'}\right)}{\hbar \omega_n} \right].
\end{align}
\end{subequations}
By introducing the relation $\mathbf{B}_n(\mathbf{r})=\frac{1}{i\omega_n}\nabla\times\mathbf{E}_n(\mathbf{r})$ and considering the definitions of the left-, right- and two-side curl, the diamagnetic tensor field and the effective couplings take the final form shown in the main text, \autoref{eq:Omega_tilde} and \autoref{eq:effcoupling}, respectively.

\section{\label{app:B} Frequency behavior of the Green's tensor}

The summation over all the modes in the effective couplings is performed using the macroscopic QED formalism. As shown in the main text, the sums transform into integrals over the full real frequency axis of functions of the EM Green's tensor $\mathbf{G}(\mathbf{r},\mathbf{r'},\omega)$. Particularly, these functions are [see \autoref{eq:effcoupling_mqed}]: $\omega\mathbf{G}(\mathbf{r},\mathbf{r'},\omega)$ in $\boldsymbol{\lambda}^{ee}\left(\mathbf{r},\mathbf{r'}\right)$, $\mathbf{G}(\mathbf{r},\mathbf{r'},\omega)\times\overleftarrow\nabla_{r'}$ in $\boldsymbol{\lambda}^{em}\left(\mathbf{r},\mathbf{r'}\right)$, $\overrightarrow\nabla_r\times\mathbf{G}(\mathbf{r},\mathbf{r'},\omega)$ in $\boldsymbol{\lambda}^{me}\left(\mathbf{r},\mathbf{r'}\right)$, $\overrightarrow\nabla_r\times\frac{1}{\omega}\mathbf{G}(\mathbf{r},\mathbf{r'},\omega)\times\overleftarrow\nabla_{r'}$ in $\boldsymbol{\lambda}^{mm}\left(\mathbf{r},\mathbf{r'}\right)$, and again the same four functions subject to the replacement $\mathbf{G}(\mathbf{r},\mathbf{r'},\omega)\rightarrow\mathbf{G}^{\textup{T}}(\mathbf{r'},\mathbf{r},\omega)$.

The Green's tensor is a causal function, which means that it is an analytic function in the upper half of the complex frequency plane $(\textup{Im}\hspace{0.05cm}\omega>0)$, with the only possibility of singularities on the real axis. Thus, knowing the pole structure on the real axis and the asymptotic high frequency behavior of $\mathbf{G}(\mathbf{r},\mathbf{r'},\omega)$ allows us to use contour integration to perform the different integrals.

In order to determine the pole structure and high frequency behavior of the Green's tensor, we follow the procedure in Ref.~\citenum{Dung2003} and start writing the nonlocal version of the Helmholtz differential equation in operator form:
\begin{equation}\label{eq:diff_eq_operator}
    \hat{\mathbf{H}}(\omega)\cdot\hat{\mathbf{G}}(\omega) = \hat{\mathbf{I}},
\end{equation}
with $\hat{\mathbf{H}}(\omega)$ being the Helmholtz operator in general nonlocal and nonreciprocal optical environments\cite{Buhmann2012}
\begin{multline} \label{eq:HelmholtzOperator}
    \hat{\mathbf{H}}(\omega) = -\hat{\mathbf{p}}\times\hat{\boldsymbol{\mu}}^{-1}(\omega)\cdot\hat{\mathbf{p}}\times - \frac{\omega}{c}\hat{\boldsymbol{\xi}}(\omega)\cdot\hat{\boldsymbol{\mu}}^{-1}(\omega)\cdot\hat{\mathbf{p}}\times
    \\
     + \frac{\omega}{c}\hat{\mathbf{p}}\times\hat{\boldsymbol{\mu}}^{-1}(\omega)\cdot\hat{\boldsymbol{\zeta}}(\omega)-\frac{\omega^2}{c^2}\left[\hat{\boldsymbol{\varepsilon}}(\omega)-\hat{\boldsymbol{\xi}}(\omega)\cdot\hat{\boldsymbol{\mu}}^{-1}(\omega)\cdot\hat{\boldsymbol{\zeta}}(\omega)\right],
\end{multline}
where $\boldsymbol{\varepsilon}(\mathbf{r},\mathbf{r'},\omega)$ is its permittivity, $\boldsymbol{\mu}(\mathbf{r},\mathbf{r'},\omega)$ is its permeability, and $\boldsymbol{\xi}(\mathbf{r},\mathbf{r'},\omega)$ and $\boldsymbol{\zeta}(\mathbf{r},\mathbf{r'},\omega)$ are its cross-magnetoelectric susceptibilities. Here, we used $\bra{\mathbf{r}}\hat{\mathbf{O}}(\omega)\ket{\mathbf{r'}}=\mathbf{O}(\mathbf{r},\mathbf{r'},\omega)$ for general tensors $\mathbf{O}$ and defined the momentum operator $\bra{\mathbf{r}}\hat{\mathbf{p}}\ket{\mathbf{r'}}=-i\overrightarrow\nabla_r\delta(\mathbf{r}-\mathbf{r'})$.
From now, we do not explicitly write the frequency dependence of the operators for simplicity. 
Decomposing the formal solution $\hat{\mathbf{G}} = \hat{\mathbf{H}}^{-1}$ of \autoref{eq:diff_eq_operator} using the longitudinal and transverse projection operators, $\hat{\mathbf{I}}^{\parallel}$ and $\hat{\mathbf{I}}^{\perp}$, we get~\cite{Dung2003}
\begin{multline}\label{eq:Feshbash_formula}
    \hat{\mathbf{G}} = \hat{\mathbf{I}}^{\parallel}\left(\hat{\mathbf{I}}^{\parallel}\hat{\mathbf{H}}\hat{\mathbf{I}}^{\parallel}\right)^{-1}\hat{\mathbf{I}}^{\parallel}
    \\
    +\left[\hat{\mathbf{I}}^{\perp}-\hat{\mathbf{I}}^{\parallel}\left(\hat{\mathbf{I}}^{\parallel}\hat{\mathbf{H}}\hat{\mathbf{I}}^{\parallel}\right)^{-1}\hat{\mathbf{I}}^{\parallel}\hat{\mathbf{H}}\hat{\mathbf{I}}^{\perp}\right]\hat{\mathbf{K}}\left[\hat{\mathbf{I}}^{\perp}-\hat{\mathbf{I}}^{\perp}\hat{\mathbf{H}}\hat{\mathbf{I}}^{\parallel}\left(\hat{\mathbf{I}}^{\parallel}\hat{\mathbf{H}}\hat{\mathbf{I}}^{\parallel}\right)^{-1}\hat{\mathbf{I}}^{\parallel}\right],
\end{multline}
where
\begin{equation}\label{eq:K}
    \hat{\mathbf{K}} = \left[\hat{\mathbf{I}}^{\perp}\hat{\mathbf{H}}\hat{\mathbf{I}}^{\perp}-\hat{\mathbf{I}}^{\perp}\hat{\mathbf{H}}\hat{\mathbf{I}}^{\parallel}\left(\hat{\mathbf{I}}^{\parallel}\hat{\mathbf{H}}\hat{\mathbf{I}}^{\parallel}\right)^{-1}\hat{\mathbf{I}}^{\parallel}\hat{\mathbf{H}}\hat{\mathbf{I}}^{\perp}\right]^{-1}.
\end{equation}
Using that $\hat{\mathbf{I}}^{\parallel}(\hat{\mathbf{p}}\times  \hat{\mathbf{O}}) = (\hat{\mathbf{O}} \times \hat{\mathbf{p}}) \hat{\mathbf{I}}^{\parallel} = \hat{\mathbf{p}} \times  \hat{\mathbf{I}}^{\parallel} =\mathbf{0}$ for general $\hat{\mathbf{O}}$, we find 
\begin{subequations}\label{eq:long_trans}
\begin{align} \label{eq:long_trans1}
    \hat{\mathbf{I}}^{\parallel}\hat{\mathbf{H}}\hat{\mathbf{I}}^{\parallel} &= \frac{\omega^2}{c^2}\hat{\mathbf{I}}^{\parallel}(\hat{\boldsymbol{\varepsilon}}-\hat{\boldsymbol{\xi}}\cdot\hat{\boldsymbol{\mu}}^{-1}\cdot\hat{\boldsymbol{\zeta}})\hat{\mathbf{I}}^{\parallel},
    \\
    \begin{split}
        \hat{\mathbf{I}}^{\parallel}\hat{\mathbf{H}}\hat{\mathbf{I}}^{\perp} &= - \frac{\omega}{c}\hat{\mathbf{I}}^{\parallel}\left(\hat{\boldsymbol{\xi}}\cdot\hat{\boldsymbol{\mu}}^{-1}\cdot\hat{\mathbf{p}}\times\right)\hat{\mathbf{I}}^{\perp} 
        \\
        &- \frac{\omega^2}{c^2}\hat{\mathbf{I}}^{\parallel}(\hat{\boldsymbol{\varepsilon}}-\hat{\boldsymbol{\xi}}\cdot\hat{\boldsymbol{\mu}}^{-1}\cdot\hat{\boldsymbol{\zeta}})\hat{\mathbf{I}}^{\perp},
    \end{split}
    \\
    \begin{split}
        \hat{\mathbf{I}}^{\perp}\hat{\mathbf{H}}\hat{\mathbf{I}}^{\parallel} &= \frac{\omega}{c}\hat{\mathbf{I}}^{\perp}\left(\hat{\mathbf{p}}\times\hat{\boldsymbol{\mu}}^{-1}\cdot\hat{\boldsymbol{\zeta}}\right)\hat{\mathbf{I}}^{\parallel} 
        \\
        &- \frac{\omega^2}{c^2}\hat{\mathbf{I}}^{\perp}(\hat{\boldsymbol{\varepsilon}}-\hat{\boldsymbol{\xi}}\cdot\hat{\boldsymbol{\mu}}^{-1}\cdot\hat{\boldsymbol{\zeta}})\hat{\mathbf{I}}^{\parallel}.
    \end{split}
    \end{align}
\end{subequations}
The frequency properties of the Green's tensor depend on the frequency behaviour of the linear response functions $\boldsymbol{\varepsilon}(\mathbf{r},\mathbf{r'},\omega)$, $\boldsymbol{\mu}^{-1}(\mathbf{r},\mathbf{r'},\omega)$, $\boldsymbol{\xi}(\mathbf{r},\mathbf{r'},\omega)$, and $\boldsymbol{\zeta}(\mathbf{r},\mathbf{r'},\omega)$. They are all analytic in the upper half of the complex plane, and also well-behaved on the real frequency axis (admit a Taylor expansion) at the position where the polarizations and magnetizations are defined. This implies, together with \autoref{eq:Feshbash_formula}, that the Green's tensor is also analytic in the upper complex frequency plane and on the real axis, except for a second order pole at $\omega=0$ [stemming from the purely longitudinal part in \autoref{eq:Feshbash_formula}, see \autoref{eq:long_trans1}]. We further find
\begin{subequations}\label{eq:myfunctions_1}
    \begin{align} 
    \lim_{\omega \to 0}   \omega^2 \mathbf{G}(\mathbf{r},\mathbf{r'},\omega) &= \lim_{\omega \to 0}   \omega^2 {}^\parallel\mathbf{G}^\parallel(\mathbf{r},\mathbf{r'},\omega)  < \infty,
        \\
   \lim_{\omega \to 0} \omega    \mathbf{G}^\perp (\mathbf{r},\mathbf{r'},\omega) &=    \lim_{\omega \to 0}  \omega\,  {}^\parallel \mathbf{G}^\perp (\mathbf{r},\mathbf{r'},\omega) \\ \label{eq:deriv1}
   & =\frac{d}{d\omega}\omega^2 \mathbf{G}^\perp(\mathbf{r},\mathbf{r'},\omega) |_{\omega =0} < \infty ,
        \\
 \lim_{\omega \to 0}  \omega   {}^\perp\mathbf{G}(\mathbf{r},\mathbf{r'},\omega) &= \lim_{\omega \to 0}  \omega \,   {}^\perp\mathbf{G}^\parallel(\mathbf{r},\mathbf{r'},\omega) \\ \label{eq:deriv2}
 &=\frac{d}{d\omega}\omega^2 \, {}^\perp\mathbf{G}(\mathbf{r},\mathbf{r'},\omega) |_{\omega =0} < \infty ,
        \\ \label{eq:deriv3}
\lim_{\omega \to 0} {}^\perp\mathbf{G}^\perp(\mathbf{r},\mathbf{r'},\omega)  &=\frac{1}{2} \frac{d^2}{d\omega^2}\omega^2 {}^\perp\mathbf{G}^\perp(\mathbf{r},\mathbf{r'},\omega) |_{\omega =0} < \infty .
    \end{align}
\end{subequations}
Here, $<\infty$ implies that the expressions remain finite and in Eqs.~\eqref{eq:deriv1},~\eqref{eq:deriv2}, and~\eqref{eq:deriv3} we used that $\omega^2 \mathbf{G}(\mathbf{r},\mathbf{r'},\omega) $ can be Taylor expanded around $\omega = 0$. Equation~\eqref{eq:myfunctions_1} shows that $\omega \mathbf{G}(\mathbf{r},\mathbf{r'},\omega)$, $  \overrightarrow\nabla_r\times  \mathbf{G} (\mathbf{r},\mathbf{r'},\omega)$, $\mathbf{G} (\mathbf{r},\mathbf{r'},\omega)\times\overleftarrow\nabla_{r'}$, and $\overrightarrow\nabla_r\times  [\mathbf{G} (\mathbf{r},\mathbf{r'},\omega)/\omega ]\times\overleftarrow\nabla_{r'} $ all have a first order pole at $\omega =0$. 

In the high-frequency limit $\left(\abs{\omega}\rightarrow \infty\right)$, the medium becomes transparent so that $\hat{\boldsymbol{\varepsilon}}\rightarrow \hat{\mathbf{I}}$, $\hat{\boldsymbol{\kappa}}\rightarrow \hat{\mathbf{I}}$, $\hat{\boldsymbol{\zeta}} \to \mathbf{0}$, and $\hat{\boldsymbol{\xi}} \to \mathbf{0}$. Also using that $\hat{\mathbf{I}}^{\perp}\hat{\mathbf{I}}^{\parallel}=\hat{\mathbf{I}}^{\parallel}\hat{\mathbf{I}}^{\perp} =\mathbf{0}$, we find from \autoref{eq:Feshbash_formula} 
\begin{align} \label{eq:GHigh}
  \lim_{|\omega| \to \infty} \frac{\omega^2}{c^2} \hat{\mathbf{G}} = - \hat{\mathbf{I}}.
\end{align}

\begin{figure}
\includegraphics[width=\linewidth]{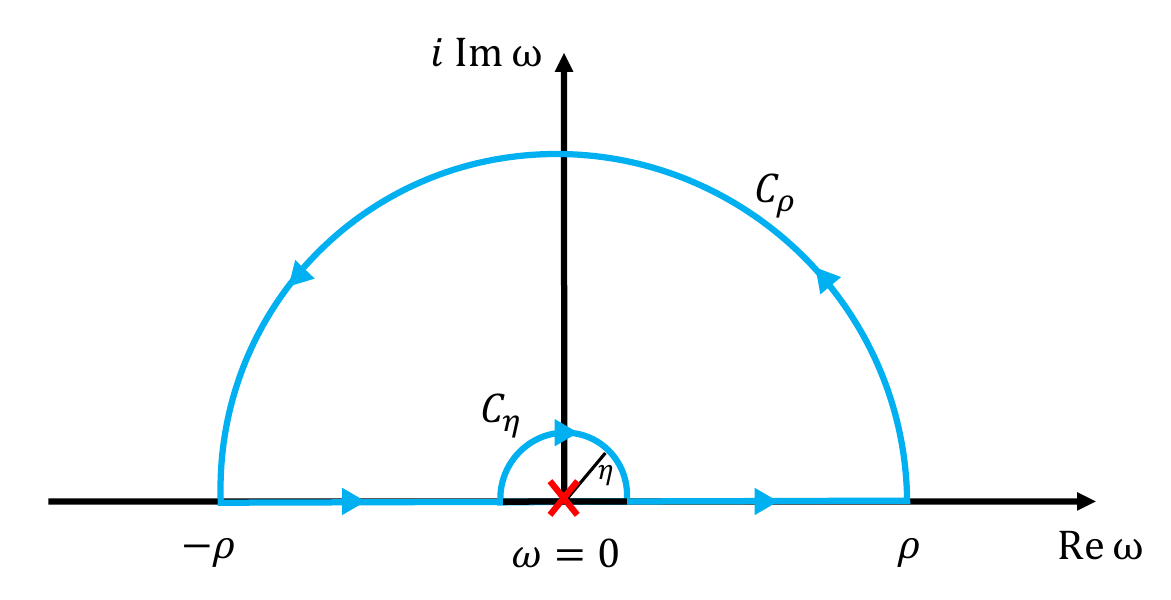}
\caption{Contour of integration $C=\left[-\rho,-\eta\right]\cup -C_{\eta}\cup\left[\eta,\rho\right]\cup C_{\rho}$ for a causal function $\chi(\omega)$ with a simple pole at $\omega=0$.}
\label{fig:n1}
\end{figure}

The above considerations show that all the integrands $\chi(\omega)$ are causal functions with simple poles at $\omega=0$. The integrals can thus be computed by constructing a semicircumference contour $C$ in the upper half of the complex plane (see \autoref{fig:n1}). The analyticity of $\chi(\omega)$ in the upper half of the complex plane implies
\begin{equation}\label{eq:Cauchy}
    \oint_{C}d\omega\hspace{0.05cm}\chi(\omega) = 0.
\end{equation}
By splitting the contour $C$ as $C=\left[-\rho,-\eta\right]\cup -C_{\eta}\cup\left[\eta,\rho\right]\cup C_{\rho}$ and taking the limit $\rho\rightarrow\infty$ and $\eta\rightarrow 0$, \autoref{eq:Cauchy} can be written as
\begin{equation}\label{eq:general_sol}
     \int_{-\infty}^{\infty}d\omega\hspace{0.05cm}\chi(\omega) = i\pi\textup{Res}\left[\chi(\omega)\right]_{\omega=0} -i\lim_{\abs{\omega}\rightarrow \infty}\int_{0}^{\pi}d\phi\hspace{0.05cm}\omega\hspace{0.05cm}\chi(\omega),
\end{equation}
where $\textup{Res}\left[\chi(\omega)\right]_{\omega=0}$ is the residue of $\chi(\omega)$ at $\omega=0$ and we substituted $\omega=\abs{\omega}e^{i\phi}$ in the last term. We can use this general formula together with \autoref{eq:myfunctions_1} and \autoref{eq:GHigh} to compute the four integrals in \autoref{eq:effcoupling_mqed}:
\begin{subequations} \label{eq:contourIntegrals}
\begin{multline}
    \int_{-\infty}^{\infty}d\omega\hspace{0.05cm}\omega\mathbf{G}(\mathbf{r},\mathbf{r'},\omega) =\\
    i\pi \lim_{\omega \to 0 }\omega^2 \mathbf{G}(\mathbf{r},\mathbf{r'},\omega) 
    - i\pi\lim_{\abs{\omega}\rightarrow \infty} \omega^2\mathbf{G}(\mathbf{r},\mathbf{r'},\omega) =\\
    i\pi\hspace{0.03cm}\lim_{\omega \to 0 }\omega^2 \mathbf{G}(\mathbf{r},\mathbf{r'},\omega)+i\pi c^2\boldsymbol{\delta}(\mathbf{r}-\mathbf{r'}),
\end{multline}\vspace{-7mm}
\begin{multline}
    \int_{-\infty}^{\infty}d\omega\hspace{0.05cm}\mathbf{G}(\mathbf{r},\mathbf{r'},\omega)\times\overleftarrow\nabla_{r'} = \\
    i\pi\hspace{0.03cm} \left[ \frac{d}{d\omega }\omega^2 \mathbf{G}(\mathbf{r},\mathbf{r'},\omega)\times\overleftarrow\nabla_{r'} \right]_{\omega =0},
\end{multline}\vspace{-7mm}
\begin{multline}
    \int_{-\infty}^{\infty}d\omega\hspace{0.05cm}\overrightarrow\nabla_r\times\mathbf{G}(\mathbf{r},\mathbf{r'},\omega) = \\
    i\pi\hspace{0.03cm} \left[\overrightarrow\nabla_r\times \lim_{\omega \to 0 }\frac{d}{d\omega }\omega^2 \mathbf{G}(\mathbf{r},\mathbf{r'},\omega)\right]_{\omega = 0} ,
\end{multline}\vspace{-7mm}
\begin{multline}
    \int_{-\infty}^{\infty}d\omega\hspace{0.05cm}\overrightarrow\nabla_r\times\frac{1}{\omega}\mathbf{G}(\mathbf{r},\mathbf{r'},\omega)\times\overleftarrow\nabla_{r'} = \\
    \frac{i\pi}{2} \left[ \overrightarrow\nabla_r\times \frac{d^2}{d\omega^2}\omega^2 \mathbf{G}(\mathbf{r},\mathbf{r'},\omega)  \times\overleftarrow\nabla_{r'} \right]_{\omega =0}.
\end{multline}
\end{subequations}
Using these integrals in \autoref{eq:effcoupling_mqed}, we recover the final expression for the effective couplings in \autoref{eq:effcoupling_mqed_2}.

\end{document}